\documentstyle{article}
\title{Quantum field theory with an interaction on the boundary}
\author{ Z. Haba\\Institute of Theoretical Physics, University of Wroclaw,
\\50-204 Wroclaw, Plac Maxa Borna 9, Poland\\e-mail:zhab@ift.uni.wroc.pl,FAX 48713214454,tel.48713759432}
\date{PACS numbers 04.62+v,02.50Cw;Key words:dimensional reduction,boundary quantum fields,
conformal field theory,functional integration}
\begin{document}
\maketitle
\begin{abstract}
We consider quantum  theory of fields $\phi$ defined on a $D$
dimensional manifold (bulk) with an interaction $V(\phi)$
concentrated on a $d<D$ dimensional surface (brane). Such a
quantum field theory can be less singular than the one in $d$
dimensions with an interaction $V(\phi)$. It is shown that scaling
properties of fields on the brane are different from the ones in
the bulk.  We discuss as an example fields on de Sitter space.
\end{abstract}
\section{ Introduction}
Models with an interaction concentrated on a $d<D$ dimensional
submanifold (brane) of a $D$ dimensional manifold (bulk) are
interesting for high energy physics as well as for statistical
physics. In the first case we consider the visible  universe as a
submanifold ( a brane \cite{gvali}\cite{sundrum})of a higher
dimensional space. Field theoretic models with an interaction on
the boundary come also from string theory \cite{horava}. In
ref.\cite{horava} the eleven dimensional gravity is interacting
with ten dimensional gauge fields living on the boundary. In
statistical physics we may consider materials with a boundary and
an interaction of some constituents placed on the boundary
\cite{lubensky}\cite{gumbs}\cite{diehl}\cite{cardy}. It is an
experimental fact \cite{dosch} that correlation functions of field
variables depending on the boundary points have critical exponents
different from the bulk correlation functions.

In this paper we discuss field theoretic models with an
interaction on the brane. We concentrate on the scalar field but
some methods and results can be generalized to gravitational and
gauge field interactions. We begin with the free propagator. We
admit any boundary condition preserving the symmetries of the free
Lagrangian. We show that a differential operator which is singular
close to the brane has the Green function which is more regular on
the brane than the one for operators with constant coefficients.
Subsequently, we discuss models with an interaction concentrated
on the brane. The functional measure is defined \cite{simon} by
its covariance (the Green function) and its mean. The mean breaks
symmetries of the  classical action. We average over  mean values
in order to preserve the symmetries. As a consequence of the more
regular behaviour of the Green function the model with an
interaction concentrated on the brane
 has milder ultraviolet divergencies.
 We give examples of nonrenormalizable theories in the bulk which
 become superrenormalizable when restricted to the boundary.
 We work mainly with the imaginary time version of quantum field theory.
 In the last section we discuss the scattering theory. The free
 particle is treated as a packet of waves on a curved
 manifold. We calculate the scattering matrix of such particles
 resulting from the $V(\phi)$ interaction concentrated
 either on the boundary or at the time $z=0$ (a kick at  a fixed
 moment).
\section{Green functions on a boundary} We
consider a $D=d+m$ dimensional Riemannian manifold of the warped
form ${\cal M}_{g}={\cal M}_{m}\times_{g}R^{d}$ \cite{warp} with
the boundary $R^{d}$ whose metric close to the boundary takes the
form
\begin{equation}
ds^{2}=G_{AB}(X)dX^{A}dX^{B}=g_{\mu\nu}(y)dx^{\mu}dx^{\nu}+g_{jk}(y)dy^{j}dy^{k}
\end{equation}
where $X=(y,x)$ are local coordinates on ${\cal M}_{g}$, $g_{jk}$
is the Riemannian metric induced on ${\cal M}_{m}$ and
$g_{\mu\nu}:{\cal M}_{m}\rightarrow R^{d^{2}} $ is a positive
definite $d\times d$ matrix function defined on ${\cal M}_{m}$.
The action for the free field $\phi$ reads
\begin{equation}
W_{0}=\int dX\sqrt{G}G^{AB}\partial_{A}\phi\partial_{B}\phi
\end{equation}
The free (Euclidean) quantum field can be defined as the one whose
propagator is determined by the Green function
\begin{equation}
 -{\cal A}{\cal G}\equiv\partial_{A}G^{AB}\sqrt{G}\partial_{B} {\cal G}=\delta
\end{equation}
where $G=\det{G_{AB}}$. In the metric (1) eq.(3) can  be expressed
as
\begin{equation}
\Big(g^{\mu\nu}(y)\sqrt{G}(y)\frac{\partial}{\partial
x^{\mu}}\frac{\partial}{\partial x^{\nu}}+\frac{\partial}{\partial
y^{j}}g^{jk}(y)\sqrt{G}(y)\frac{\partial}{\partial
y^{k}}\Big){\cal G}=\delta
\end{equation}
The solution of eq.(3) is not unique. If ${\cal G}^{\prime}$ is
another solution of eq.(3) then ${\cal G}^{\prime}={\cal G}+{\cal
R}$ where ${\cal R}$ is a  solution of the equation
\begin{equation}
{\cal A}{\cal R}=0
\end{equation}
We can determine ${\cal G}$ unambiguously imposing some additional
requirements, e.g., requiring that ${\cal G}=0$ on the boundary or
that ${\cal G} $ be scale invariant on the boundary.

 We assume that the metric is
scale invariant
\begin{equation} g_{\mu\nu}(\lambda
y)=\lambda^{2\alpha}g_{\mu\nu}(y)
\end{equation}
and
\begin{equation}g_{jk}(\lambda y)=\lambda^{2\beta}g_{jk}(y)
\end{equation}
It follows that
\begin{displaymath}
G(\lambda y)=\lambda^{2d\alpha+2m\beta}G(y)
\end{displaymath} Let us consider ${\cal G}_{\lambda,\rho}(
x, y; x^{\prime}, y^{\prime})\equiv{\cal G}(\rho x,\lambda y;\rho
x^{\prime},\lambda y^{\prime})$. It satisfies the equation
\begin{equation}
\begin{array}{l}
\lambda^{m}\rho^{d}\Big(\lambda^{-2\alpha+m\beta+d\alpha}g^{\mu\nu}(
y)\sqrt{G}( y)\frac{\partial}{\partial \rho
x^{\mu}}\frac{\partial}{\partial \rho x^{\nu}}\cr
+\lambda^{m\beta+d\alpha-2\beta}\frac{\partial}{\partial \lambda
y^{j}}g^{jk}( y)\sqrt{G}( y)\frac{\partial}{\partial \lambda
y^{k}}\Big){\cal G}=\delta(x-x^{\prime})\delta(y-y^{\prime})
\end{array}
\end{equation}
Eq.(8) is identical with eq.(4) but expressed in rescaled
coordinates (the scale invariance (6)-(7) has been applied).  Let
us choose $\rho$ such that the scale factors in the two terms in
eq.(8) are equal
\begin{equation}\lambda^{-2\alpha+m\beta+d\alpha}
\rho^{-2}= \lambda^{m\beta+d\alpha-2\beta-2} \end{equation} Hence,
\begin{equation}
\rho=\lambda^{1-\alpha+\beta}
\end{equation}
Then, it follows that\begin{equation}
\Big(g^{\mu\nu}(y)\sqrt{G}(y)\frac{\partial}{\partial
x^{\mu}}\frac{\partial}{\partial x^{\nu}}+\frac{\partial}{\partial
y^{j}}g^{jk}(y)\sqrt{G}(y)\frac{\partial}{\partial
y^{k}}\Big)\hat{{\cal G}}=\delta
\end{equation}
where
\begin{equation}
\hat{{\cal
G}}(y,x;y^{\prime},x^{\prime})=\lambda^{(D-2)(1+\beta)}{\cal
G}(\lambda y,\lambda^{1-\alpha+\beta}x;\lambda
y^{\prime},\lambda^{1-\alpha+\beta} x^{\prime})
\end{equation}
$\hat{{\cal G}}$ and ${\cal G}$ satisfy the same equation (4).
Hence, for the scale invariant solution $\hat{{\cal G}}={\cal G}$
(there are many solutions of eq.(4) but the scale invariant
solution is  unique). We are interested in the boundary Green
function ${\cal G}(0,x;0,x^{\prime})\equiv {\cal G}_{E}(\vert
x-x^{\prime}\vert)$. ${\cal G}_{E}$ depends solely on the
Euclidean distance $\vert x-x^{\prime}\vert $ as a consequence of
the translational invariance of eq.(4) and its rotational
invariance when $y=y^{\prime}=0$. Choosing in eq.(12)
\begin{equation}
\lambda=\vert x-x^{\prime}\vert^{-\frac{1}{1-\alpha+\beta}}
\end{equation}
we obtain
\begin{equation}{\cal G}_{E}(\vert x-x^{\prime}\vert)=K
\vert x - x^{\prime}\vert^{-(D-2)\sigma }
\end{equation}
where
\begin{equation}
\sigma=\frac{1+\beta}{1-\alpha+\beta}
\end{equation}
and $K(\alpha,\beta,D)={\cal G}(0,e;0.0)$ (here $e\in R^{d}$ is an
arbitrary  vector such that $\vert e\vert=1$). Some authors
\cite{arefeva}\cite{muck}\cite{fischler} in de Sitter case require
that ${\cal G}=0$ at $y=0$. In such a case $K=0$ in eq.(14) and
our result is trivial. However, in other models of QFT on  de
Sitter space
\cite{ford}\cite{schom}\cite{tsamis}\cite{allen}\cite{birrell-davis}
${\cal G} $ does not vanish at $y=0$.

 It will be
useful to rewrite eq.(4) in the momentum space. Let
\begin{equation}
{\cal G}(y,x;y^{\prime},x^{\prime})=(2\pi)^{-m}\int dp
\exp(ip(x-x^{\prime})) \tilde{{\cal G}}(p;y,y^{\prime})
\end{equation}
Then, $\tilde{{\cal G}}$ satisfies the equation
\begin{equation}
\Big(-p_{\mu}p_{\nu}g^{\mu\nu}(y)\sqrt{G}(y)+\frac{\partial}{\partial
y^{j}}g^{jk}(y)\sqrt{G}(y)\frac{\partial}{\partial
y^{k}}\Big)\tilde{{\cal G}}=\delta(y-y^{\prime})
\end{equation}
Let $\tilde{{\cal G}}_{\lambda,\gamma}(p;y,y^{\prime})=
\tilde{{\cal G}}(\gamma p;\lambda y,\lambda y^{\prime})$. Then,
$\tilde{{\cal G}}_{\lambda,\gamma}$ is a solution of the equation
\begin{equation}
\begin{array}{l}
\lambda^{m}\Big(\lambda^{-2\alpha+m\beta+d\alpha}\gamma^{2}p_{\mu}p_{\nu}g^{\mu\nu}(
y)\sqrt{G}( y)\cr +\lambda^{-2\beta
+m\beta+d\alpha}\frac{\partial}{\partial \lambda y^{j}}g^{jk}(
y)\sqrt{G}( y)\frac{\partial}{\partial \lambda
y^{k}}\Big)\tilde{{\cal G}}_{\lambda,\gamma}=\delta(y-y^{\prime})
\end{array}
\end{equation}
We choose
\begin{displaymath}
\gamma=\lambda^{-1-\beta+\alpha}
\end{displaymath}
Then, we obtain (similarly as in eq.(12)) from the uniqueness of
the solution of eq.(17)
\begin{equation}
\tilde{{\cal G}}(p; y, y^{\prime}) =\lambda^{m-2-2\beta
+m\beta+d\alpha}  \tilde{{\cal
G}}(\lambda^{-1-\beta+\alpha}p;\lambda y,\lambda y^{\prime})
\end{equation}
 Setting
\begin{displaymath}
\lambda=\vert p\vert^{\frac{1}{1-\alpha+\beta}}
\end{displaymath}
and $y=y^{\prime}=0$ in eq.(19) leads to the result
\begin{equation}
 \tilde{{\cal G}}(p;0,0)=\tilde{K}\vert p\vert^{-2\omega}
 \end{equation}
 where
 \begin{equation}
 -2\omega=\frac{(m-2)(1+\beta)+d\alpha}{1+\beta-\alpha}
 \end{equation}
and $\tilde{K}(\alpha,\beta,D)=\tilde{{\cal G}}(\tilde{e};0,0) $ (
$\tilde{e}$ is an arbitrary unit vector in $R^{d}$).

 As an example of an application of
eqs.(20)-(21) we could consider de Sitter space with the metric
(it is of the type (1);this is Euclidean AdS$_{D}$ \cite{witten},
the notion of the boundary in AdS and its relation to Euclidean
AdS is discussed in \cite{maldacena} \cite{witten})
\begin{equation}
 ds^{2}=dt^{2}+\exp(2Ht) (dx_{1}^{2}
 +....+dx_{D-1}^{2})=y^{-2}(dy^{2}+ dx_{1}^{2}
 +....+dx_{D-1}^{2})
 \end{equation}
where $y=H^{-1}\exp(-Ht)$ .

 Then, from eqs.(20)-(21)\begin{equation}
 \tilde{{\cal G}}(p;0,0)=\tilde{K}\vert p\vert^{-D+1}
 \end{equation}
 The Fourier transform of $\tilde{{\cal G}}$ in eq.(23) is infrared
 divergent. It can be defined as a distribution on a set of
 functions
 vanishing at  $p=0$. Then,
 \begin{equation}
 {\cal G}_{E}(\vert x-x^{\prime}\vert)=-K\ln\vert
 x-x^{\prime}\vert
 \end{equation}Eq.(24) gives the form of the Green function
 for $\sigma=0$ in eq.(14).

 Generalizing eq.(4) we could consider a system of equations
  for a tensorial
Green function ${\cal G}_{\Gamma\Omega}$
\begin{equation}
\Big(a_{\Gamma\Sigma}^{\mu\nu}(y)\frac{\partial}{\partial
x^{\mu}}\frac{\partial}{\partial x^{\nu}}+\frac{\partial}{\partial
y^{j}}b_{\Gamma\Sigma}^{jk}(y)\frac{\partial}{\partial
y^{k}}+A_{\Gamma\Sigma}^{\mu}\frac{\partial}{\partial
x^{\mu}}+B_{\Gamma\Sigma}^{j}\frac{\partial}{\partial
y^{j}}+V_{\Gamma\Sigma}(y)\Big)\hat{{\cal
G}}_{\Sigma\Omega}=\delta_{\Gamma\Omega}
\end{equation}
here a sum over repeated indices $\Sigma$ is assumed, the
$\delta_{\Gamma\Omega}$ function is the product of the usual
$\delta$ function of eq.(8) and a scale invariant tensor with
indices $\Gamma\Omega$. We assume that the coefficients have the
following scaling properties
\begin{equation}
a^{\mu\nu}(\lambda y)=\lambda^{(d-2)\alpha+m\beta}a^{\mu\nu}(y)
\end{equation}
\begin{equation} b^{jk}(\lambda y)
=\lambda^{d\alpha+(m-2)\beta}b^{jk}(y)
\end{equation}
\begin{equation}
A^{\mu}(\lambda y)=\lambda^{(m-2)\beta+d\alpha-1}A(y)
\end{equation}
\begin{equation}
B^{j}(\lambda y)=\lambda^{(m-1)\beta+(d-1)\alpha-1}B^{j}(y)
\end{equation}
and
\begin{displaymath}
V(\lambda y)=\lambda^{(m-2)\beta+d\alpha-2}V(y)
\end{displaymath}
Then,  repeating our scaling arguments of this section we could
derive the results (14)-(15) for ${\cal G}_{\Gamma\Omega}$ and
(20)-(21) for its Fourier transform. Such results may be
applicable to propagators describing an interaction of gravity on
the bulk and on the brane with gauge fields on the brane as in
\cite{horava}\cite{gvali}.

\section{DeWitt expansion }
Let us discuss now scaling properties of the Green functions from
the point of view of DeWitt expansion \cite{dewitt} and the
Hadamard representation of the Green functions \cite{hadamard}.
According to the DeWitt suggestion we can solve the equation (for
a non-negative operator ${\cal A}$)
\begin{equation}
-\cal{A}{\cal G}=\delta
\end{equation}
by means of the heat kernel which is the fundamental solution of
the heat equation
\begin{equation}
-\frac{d}{d\tau}{\cal K}_{\tau}={\cal A}{\cal K}_{\tau}
\end{equation}
Namely,
\begin{equation}
{\cal G}=\int_{0}^{\infty}d\tau {\cal K}_{\tau}
\end{equation}
Then,  DeWitt makes the assumption
\begin{equation}
{\cal
K}_{\tau}(X,X^{\prime})=(2\pi\tau)^{-\frac{D}{2}}\exp(-\frac{\sigma(X,X^{\prime})}{2\tau})
\Lambda(\tau,X,X^{\prime})
\end{equation}
where $\sigma(X,X^{\prime})$ is the square of the geodesic
distance between $X$ and $X^{\prime}$ . $\Lambda$ has a Taylor
expansion in powers of $\tau$ for a manifold without boundary
(powers of $\sqrt{\tau}$ may appear if the manifold has a
boundary) . Performing the integral over $\tau$ in eq.(32) we
obtain the short distance expansion of ${\cal G}(X,X^{\prime})$.

     We do not know the formula for $\sigma$ in general.
     We can explain  the expansion (33) in the hyperbolic case (22).
     Then,
     \begin{equation}
    ch( \sigma(X,X^{\prime}))=1+(2yy^{\prime})^{-1}((x-x^{\prime})^{2}+(y-y^{\prime})^{2})
     \end{equation}
In the hyperbolic space the heat kernel is known exactly
\cite{kruczenski}. The DeWitt  expansion holds true ($\Lambda$ has
an expansion in powers of $ \sigma$; no $\tau$ dependence). We can
see that

\begin{equation}
\sigma(X,X^{\prime})^{-1}\rightarrow 0
\end{equation}
when $y\rightarrow 0$  ($t\rightarrow \infty $ in eq.(22))and $x
\neq x^{\prime}$. It follows from eq.(32) that if $D=2k+1$ is odd
then
\begin{equation}
{\cal G}_{E}(x-x^{\prime})=0
\end{equation}
and if $D=2k$ is even then
\begin{equation}
{\cal G}_{E}(x-x^{\prime})=-K\ln \vert x-x^{\prime}\vert
\end{equation}
The conclusion (37) coincides with eq.(24).

\section{Interacting fields on the boundary} In the model (2) we consider an interaction
$V$ concentrated on the boundary $y=0$ (we do not treat here a
coordinate independent geometric description of the boundary but
restrict ourselves to the model (1))
\begin{equation} W=W_{0}+W_{I}=\int
dX\sqrt{G}G^{AB}\partial_{A}\phi\partial_{B}\phi +\int dX\sqrt{G}
\delta(y)V(\phi)
\end{equation}
We define the functional measure
\begin{equation}
d\mu(\phi)=Z^{-1}{\cal D}\phi\exp(-W)\equiv
Z^{-1}d\mu_{0}(\phi)\exp(-W_{I})
\end{equation}
where the Gaussian measure is
\begin{displaymath} d\mu_{0}(\phi) ={\cal
D}\phi\exp(-W_{0})\end{displaymath} The partition function
\begin{displaymath}
Z=\int d\mu_{0}\exp(-W_{I}) \end{displaymath} determines a
normalization factor. The Gaussian measure is defined \cite{simon}
by the mean
\begin{displaymath}
\int d\mu_{0}(\phi)\phi(X)\equiv \langle \phi(X)\rangle
\end{displaymath}and the covariance
\begin{equation}
{\cal G}(X,X^{\prime})=\int
d\mu_{0}(\phi)(\phi(X)-\langle\phi(X)\rangle)(\phi(X^{\prime})-
\langle\phi(X^{\prime})\rangle)
\end{equation}
In the papers on AdS-CFT correspondence
\cite{witten}\cite{polyakov}\cite{maldacena}\cite{muck}\cite{arefeva}
the choice is made ${\cal G}(X,X^{\prime})=G_{D}(X,X^{\prime})$
where $G_{D}$ is the Dirichlet Green function (vanishing on the
boundary) and $\langle \phi(x)\rangle=\psi(x)$ where $\psi(x)$ is
the solution of
 the equation
\begin{displaymath}
{\cal A}\psi=0
\end{displaymath}
 with a fixed boundary condition.  However,
the choice of the boundary field $\psi\neq 0$ breaks the
rotational and translational invariance in the $x$ variables
present in the classical action (38)with the metric (1). The
approach with the classical boundary field and the Dirichlet
boundary condition leads to a different quantum field theory than
the one developed in
refs.\cite{ford}\cite{birrell-davis}\cite{schom}\cite{tsamis}\cite{guth}
(see
also \cite{allen}\cite{dolgov}). In our approach the QFT is
determined by the choice of the Green function ${\cal G}$ (40) (
we set the mean $\langle \phi\rangle =0$).
 We choose the Green function ${\cal G}$ which has the symmetries
 of the action $W_{0}$ (2). Hence, the functional measure (39)
 will have the symmetries of the action (38).

 We can show that our approach is equivalent to a quantization
 with a given classical solution $\psi$ if subsequently an average over
 all such solutions is performed.
 We assume that in the sense  of bilinear forms
\begin{equation}
{\cal G}\geq G_{D}
\end{equation}
Such an inequality follows from the maximum principle for elliptic
operators \cite{jaffe}\cite{gilbarg}. The inequality (41) holds
true also in the non-elliptic cases discussed in our earlier
papers \cite{haba1}\cite{haba2}. Using eq.(41) we may write
\begin{displaymath} {\cal G}(X,X^{\prime})=
G_{D}(X,X^{\prime})+{\cal G}_{B}(X,X^{\prime})
\end{displaymath}
where ${\cal G}_{B}$ is a non-negative bilinear form and ${\cal
G}_{B}(0,x;0,x^{\prime})={\cal G}_{E}(x-x^{\prime})$. Then,
\cite{simon}\cite{guerra}\begin{displaymath} \int
d\mu_{0}(\phi)\exp(-W_{I}(\phi)F(\phi)=\int d\mu_{D}(\phi_{D})
d\mu_{B}(\phi_{B})\exp(-W_{I}(\phi_{D}+\phi_{B}))F(\phi_{D}+\phi_{B})
\end{displaymath}
where $\phi_{D}$ is a random field with the covariance $G_{D}$ and
$\phi_{B}$ is the random field with the covariance ${\cal G}_{B}$.
Let us note that because $G_{D}$ as well as ${\cal G}$ satisfy the
same equation (2) then their difference satisfies the equation
\begin{equation}
{\cal A}(X){\cal G}_{B}(X,X^{\prime})={\cal A}(X^{\prime}){\cal
G}_{B}(X,X^{\prime})=0
\end{equation}
We can solve eq.(42) with the given boundary condition ${\cal
G}_{E}$
\begin{equation}
{\cal G}_{B}(X,X^{\prime})=\int dx_{b}\sqrt{g}\int
dx_{b}^{\prime}\sqrt{g}{\cal D}(X,x_{b}) {\cal
D}(X^{\prime},x_{b}^{\prime}){\cal G}_{E}(x_{b}-x_{b}^{\prime})
\end{equation}
where ${\cal D}$ is the Green function solving the Dirichlet
problem (the boundary to bulk propagator). It follows that
\begin{equation}
\phi_{B}(X)=\int dx_{b}\sqrt{g}{\cal D}(X,x_{b})\Phi(x_{b})
\end{equation}
where $\Phi$ is the Gaussian random field defined on the boundary
with the covariance
\begin{equation}
\langle \Phi(x)\Phi(x^{\prime})\rangle ={\cal G}_{E}(x-x^{\prime})
\end{equation}
 In the case of the de Sitter space the solution
 of the Dirichlet boundary problem (44) can be expressed
 in the form
 \begin{equation}
 \phi(X)=y^{\frac{d}{2}}\int dp\exp(ipx)
 \vert p\vert^{\frac{d}{2}}K_{\frac{d}{2}}(\vert p\vert
 y)\tilde{\Phi}(p)
 \end{equation}
 where $\tilde{\Phi}$ is the Fourier transform of $\Phi$
 and $K_{\nu}$ is the modified Bessel function of order $\nu$. From
 eq.(23)\cite{haba1}
 \begin{displaymath}
 \langle \tilde{\Phi}(p)\tilde{\Phi}^{*}(p^{\prime})\rangle=
 \delta(p-p^{\prime})\vert p\vert^{-d}
 \end{displaymath}

 The Schwinger functions
are defined as moments of the measure $\mu$
\begin{displaymath} \langle\phi(X_{1})....\phi(X_{k})\rangle=Z^{-1} \int d\mu(\phi) \phi(X_{1}).......\phi(X_{k})
\end{displaymath}
We calculate these Schwinger functions in the $N$-th order of the
perturbation expansion

\begin{equation}
\begin{array}{l}
\langle\phi(X_{1})....\phi(X_{k})\rangle_{N}=\cr
 Z^{-1}
\int
dX_{1}^{\prime}....dX_{N}^{\prime}\delta(y_{1}^{\prime})....\delta(y_{N}^{\prime})\cr
\int d\mu_{0} \phi(X_{1}).......\phi(X_{k})
V(\phi(X_{1}^{\prime})).....V(\phi(X_{N}^{\prime}))
\end{array}\end{equation}
If $V(\phi)$ is a normal-ordered polynomial of order $r$ then the
Schwinger functions of  order $N$ are expressed by a product of at
most $k$ Green functions ${\cal G}(y,x;0,x^{\prime})$ and at most
$(rN)!$ Green functions ${\cal
G}(0,x_{j}^{\prime};0,x_{k}^{\prime})$. It follows that if the
Green functions at $y=y^{\prime}=0$ are sufficiently regular
(depending on $\alpha$ and $\beta$ in eqs.(14)-(15)) and the
integration over $x$ in the interaction $W_{I}$ is restricted to a
finite volume $\Lambda$ then the Schwinger functions (47) for
non-coinciding points $X_{j}=(0,x_{j})$ are finite and non-zero .
In fact, the amputated Schwinger functions (when the propagators
corresponding to the lines connecting the external points are
removed) coincide with the ones for the $V(\phi)$ theory in
$R^{d}$ calculated with the propagator (14). If in
eq.(47)$X_{j}=(y_{j},x_{j})$ and we set all $y_{j}=0$ then the
resulting perturbative quantum field theory coincides with the
$V(\phi)$ theory where the conventional propagators are replaced
with ${\cal G}_{E}$ of eq.(14). In particular, the partition
function Z in eq.(40) can be finite and non-zero even though the
bulk theory is non-renormalizable. Note, that this result is a
consequence of the  singularity of the metric on the boundary.
Adding the boundary interaction $W_{I}$ to the regular free part
$W_{0}$
 would lead to the theory with the same singularity as the free
 field theory in $d+1$ dimensions.

Let us write the  propagator (14) in the form
\begin{equation}
{\cal G}_{E}=K \vert   x-  x^{\prime}\vert^{-d+\rho}
\end{equation}then
the effective field theory on the boundary $y=0$ could be
represented by an equivalent functional measure $\hat{\mu}$ of the
form (39) with
\begin{equation}
\hat{W}=\hat{W}_{0}+\hat{W}_{I}=c_{0}\int dx\phi
(-\triangle)^{\frac{\rho}{2}}\phi+\int dxV(\phi)
\end{equation}where $\triangle$ is the Euclidean $d$-dimensional
Laplacian. If the theory (49) is to be scale invariant (up to the
log-terms coming from the renormalization)
\begin{equation}
\phi( x)\simeq \lambda^{\nu}\phi(\lambda  x)
\end{equation}
then from the propagator (48) it follows that \begin{equation}
d-\rho=2\nu\end{equation} Hence, the interaction
\begin{equation}
\int dxV(\phi)=\kappa\int dx\phi^{r}
\end{equation}
is scale invariant if
  the order $r$ of the interaction is
related to $\nu$ (and by eq.(51) to $\rho$)
\begin{equation}
\nu=\frac{d}{r}
\end{equation}
From eqs.(51) and (53) it follows that
\begin{equation}
\rho=d(1-\frac{2}{r})
\end{equation}
If $\rho=d$ then  from eq.(54) we obtain $r=\infty$. This case
corresponds to the exponential potential $V(\phi)=\kappa
\exp(\phi)$. In general, we obtain  simple fractional scaling
dimensions (54) determined uniquely by the natural numbers $d$ and
$r$. From eq.(15) and eq.(54)
\begin{equation}
 2\nu=(d-1)\frac{1+\beta}{1-\alpha+\beta}
\end{equation}
Hence, the geometry of the surface imposes conditions on the
scaling exponents and the form of the scale invariant interaction.
From eq.(53) $r=\frac{d}{\nu}$ hence the order of the scale
invariant interaction is also determined by the surface geometry
\begin{equation}
r=\frac{2d(1-\alpha+\beta)}{(d-1)(1+\beta)} \end{equation} If
$\beta=-1$ then $\nu=0$ and $\rho=d$. In such a case from the
scale invariant fields (24) (of dimension zero) we can form
conformal fields of higher dimensions by means of exponential
functions in a similar way as in two dimensional conformal field
theory \cite{furlan-petkova} .

Let us note that the theory with an interaction on the boundary
can be considered as a scaling limit of  the one with an
interaction in the bulk. For this purpose define
$V_{\lambda}(\phi(y, x))= V(\phi(\lambda y,x))$, calculate the
Schwinger functions (47) perturbatively and at the end take the
limit $\lambda\rightarrow 0$.

\section{${\cal M}^{d+1}\rightarrow R^{d} $ reduction}
If $m=1$ then we may change coordinates in eq.(4) introducing a
new coordinate $z$ instead of $y$ in such a way that
\begin{equation} \frac{dy}{dz}=g^{DD}(y)\sqrt{G}(y)
\end{equation}
Then,  eq.(4) reads
\begin{equation}
\Big(g^{DD}(y)G(y)g^{\mu\nu}(y)\frac{\partial}{\partial
x^{\mu}}\frac{\partial}{\partial x^{\nu}}+\partial_{z}^{2}
\Big){\cal
G}=g^{DD}\sqrt{G}\delta(y-y^{\prime})\delta(x-x^{\prime})=\delta(z-z^{\prime})\delta(x-x^{\prime})
\end{equation}
The action (38) takes the form
\begin{equation}
\begin{array}{l}
W=\int dydx\sqrt{G}\Big(g^{DD}\partial_{y}\phi\partial_{y}\phi
+g^{\mu\nu}\partial_{\mu}\phi\partial_{\nu}\phi
+\kappa_{0}\delta(y)V(\phi)\Big) \cr =\int dzdx
\partial_{z}\phi\partial_{z}\phi+\int dzdxg^{DD}G
g^{\mu\nu}\partial_{\mu}\phi\partial_{\nu}\phi +\kappa_{0}\int
dzdx\sqrt{G}\delta(z)V(\phi)
\end{array}
\end{equation}
In scale invariant models $G(0)$ is either zero or infinite. Then,
we must renormalize the interaction defining
$\hat{\kappa}_{0}=\kappa_{0}\sqrt{G}(0)$.
  We have studied
the behaviour of the Green functions ${\cal G}$ of eq.(58) in
refs.\cite{haba1}\cite{haba2}. Assuming that
\begin{equation}
g^{\mu\nu}g^{DD}G\simeq \delta^{\mu\nu} \vert z\vert^{2\gamma}
\end{equation}
for a small $z$ we have shown (this is the same result as the one
in eq.(14)) that for small $\vert x - x^{\prime}\vert $
\begin{equation}
{\cal G}_{E}( x- x^{\prime})\simeq \vert  x-
x^{\prime}\vert^{-d+\frac{1}{1+\gamma}}
\end{equation}
As discussed in sec.4 the theory at $z=0$ is the same as the one
with the  interaction $V$ and the propagator (48). Hence,
depending on the value of $\gamma \geq -1+\frac{1}{d}$ the
boundary field theory can be much more regular than the bulk field
theory . When $\gamma=-1 +\frac{1}{d}$ ( as in the case of de
Sitter space) then the model (59) becomes superrenormalizable for
polynomial interactions.

 It is
interesting to consider $d=2$ and $\rho=2$ with the exponential
interaction $\exp \phi $ on the boundary. Then, the boundary
correlation functions are conformal invariant (the Liouville model
\cite{liouville}) whereas the bulk correlation functions  are not
renormalizable.

The other interesting case is the boundary of the Euclidean
(anti)de Sitter space \cite{witten}(eqs.(22)-(24))with an
exponential interaction. Again the boundary correlation functions
of exponentials can be conformal invariant \cite{furlan-petkova}
whereas the bulk correlation functions are non-renormalizable.

\section{Semi-infinite statistical systems} In this section
we relate the model of sec.4 to some models of statistical physics
\cite{lubensky} \cite{gumbs}\cite{diehl}. Let us consider
Ginzburg-Landau Lagrangians with a scalar field $\phi(X)$ where
$X\in R^{D}$, $X=(z, x)$ with $ x\in R^{d}$ and $z>0$. The
boundary is at $z=0$. We consider  the action
\begin{equation}
\begin{array}{l}
W(\phi)=\int dz \int
dx\Big(c_{1}\sum_{\mu=1}^{d}\partial_{\mu}\phi
\partial_{\mu}\phi+
c_{0}\delta(z)\sum_{\mu=1}^{d}\partial_{\mu}\phi
\partial_{\mu}\phi+c_{3}\partial_{z}\phi\partial_{z}\phi
\cr
+\kappa_{1}V_{1}(\phi) +\kappa_{0}\delta(z)V_{0}(\phi)\Big)
\end{array}
\end{equation}
The free propagator for a perturbation expansion around the
Gaussian theory is determined by the equation
\begin{equation}
\Big(c_{1}\sum_{\mu=1}^{d}\partial_{\mu}
\partial_{\mu}+
c_{0}\delta(z)\sum_{\mu=1}^{d}\partial_{\mu}
\partial_{\mu}+c_{3}\partial_{z}\partial_{z}\Big ){\cal G}=\delta
\end{equation}
This model is related to the models of secs.4 and 5. We have
studied the propagator (63) in refs.\cite{haba1}\cite{haba2}. If
in the equation (59)
$a^{\mu\nu}=g^{\mu\nu}g^{DD}G=\delta^{\mu\nu}\vert z\vert^{-1}$
then the  scaling properties  of $a^{\mu\nu}$ are the same as the
ones of $\delta(z)$ (i.e., $\delta(\lambda z)=\lambda^{-1}\delta
(z)$) . In such a case the propagator (63) has the same short
distance behaviour as the one for
$a^{\mu\nu}(y)=\delta^{\mu\nu}\vert y\vert ^{-1}$ and $b=1$ in
eqs.(26)-(27) ( if $c_{1}=0$ then these propagators coincide).
Now, in eq.(26) $(d-2)\alpha+m\beta=-1$ and
$(m-2)\beta+d\alpha=0$. Hence, in eq.(14) $\sigma(d-1)=d-2$ and if
$c_{1}=0$ then we have exactly

 \begin{equation}
 {\cal G}_{E}(\vert x-x^{\prime}\vert)=K\vert
 x-x^{\prime}\vert^{-d+2}
\end{equation}
As a consequence the propagator in $ (d+1)$ dimensions behaves as
the one in $d$ dimensions. In particular, if $d+1=3$ then with the
gradient term on the boundary  ($c_{0}\neq 0$ )we obtain the
logarithmic short distance behaviour of boundary correlation
functions. In such a case the model on the boundary
($\kappa_{1}=0$) has the properties of the two-dimensional
Euclidean field theory. If $V_{0}(\phi)=\phi^{6}$ then we obtain a
renormalizable field theory in the bulk which on the boundary
reduces to the well-studied model of a tricritical phase
transition . If $d+1=4$ and $V_{0}(\phi)=\phi^{4}$ then we have a
typical superrenormalizable Landau-Ginzburg model. The model is
superrenormalizable because the truncated Green functions are the
same as in the threedimensional theory. The model in $d+1=5$ with
the $\phi^{4}$ interaction on the boundary is still
renormalizable. In terms of the lattice approximation our
semi-infinite model with $c_{1}=0$ describes spins whose
interaction  in the bulk is restricted to the lines perpendicular
to the boundary. It is still surprising that although the system
is $D$ dimensional the correlation functions behave like the ones
of an $d$-dimensional system in spite of the opportunity for the
interaction to spread into the $D$-th dimension.

\section{Scattering theory with an interaction concentrated on the
boundary}
 We could approach the models of secs.4-6 in the conventional Hamiltonian
 framework. We consider the scattering theory for the  field equations
\begin{equation}
{\cal A}\phi=-\delta(z)V^{\prime}(\phi)
\end{equation}
We could treat $z$ either as a spatial coordinate or as a time.
Let us concentrate here on the latter interpretation. Then, for
$z\rightarrow - \infty $ the interaction is switched off and
$\phi\rightarrow \phi^{in}$ where
\begin{equation} {\cal
A}\phi^{in}=0
\end{equation}
We consider here the real time $z$ and the wave operator ${\cal
A}=-\partial_{z}^{2}-A^{2}$ where $A^{2}$ is a positive operator.
We quantize the field $\phi^{in}$ and construct the interaction
Hamiltonian $H_{I}(z)= \delta(z)\int dx V(\phi^{in})$. We can
derive the S-matrix describing an interaction $V(\phi)$ of the
field $\phi^{in}$ by means of the  conventional reduction formulas
\cite{birrell-davis} \cite{birrell-taylor}. We assume that the
geometry is fixed and therefore the asymptotic states are
constructed on a given gravitational background. As usual the
S-matrix is determined by the time-ordered correlation functions
\begin{equation} \langle T\Big(\phi(X_{1})....\phi(X_{n})\Big)\rangle
=\langle T\Big(\phi^{in}(X_{1})... \phi^{in}(X_{n})\exp(-i\int dzd
x \delta(z)\sqrt{g}V(\phi^{in}))\Big)\rangle
\end{equation}
The expectation value on the rhs is calculated in the vacuum for
the free field $\phi^{in}$ defined on the curved manifold (there
may be many vacuum states and it belongs to the theory to choose
one of them). In perturbation theory (in the Euclidean region)the
correlation functions are expressed by the Green functions (47).
The S-matrix is determined by $\tau$-functions
\begin{displaymath}
S=1+\sum_{n}\frac{(-i)^{n}}{n!}\int
dX_{1}....dX_{n}\tau(X_{1}....X_{n})
:\phi^{in}(X_{1}).....\phi^{in}(X_{n}):+ ...
\end{displaymath}
where  \begin{displaymath}
 \tau(X_{1},...,X_{n})={\cal A}(X_{1})....{\cal A}(X_{n})
 \langle T\Big(\phi(X_{1})...\phi(X_{n})\Big)\rangle
 \end{displaymath}
 Hence, from eq.(47) the ${\cal A}$ operators cancel the external propagators
 and the $\tau$ functions depend  on Green functions defined at $z=0$.

  Let us consider some examples. First, the case when in eq.(58)
  \begin{equation}
  g^{DD}Gg^{\mu\nu}(y)=\delta^{\mu\nu}\vert z\vert^{-1}
  \end{equation}
  (this is an analogue of eq.(63)).
  Then
  \begin{equation}
  {\cal A}=\vert z\vert^{-1}\triangle -\partial_{z}^{2}
  \end{equation}
  The solutions of eq.(69) have the form
  \begin{equation}
  \phi^{in}(z, x)=\sqrt{z}\int d p\Big( a( p)H_{1}^{(1)}(C\vert p\vert \sqrt{z})+ a^{+}(
  p)\overline{H_{1}^{(1)}}(C\vert p\vert \sqrt{z})\Big)\exp(i px)
  \end{equation}
where $H_{\nu}^{(1)}$ is the Hankel function of order $\nu$
\cite{stegun} and $C$ is a positive constant.

   A
  quantization of $a$ and $a^{+}$ (as creation and annihilation operators)
  leads to a quantum scattering theory described by
  an expansion of the S-matrix in asymptotic fields $\phi^{in}$.
  Note, that at $z=0$ the Feynman propagator
  \begin{equation}
  {\cal G}_{E}( x- x^{\prime})=K\int d p\exp(i
  p( x- x^{\prime}))\vert p\vert^{-2}
  \end{equation}
  for ${\cal A}$ in eq.(69) is equal to
   the one for a  massless Euclidean  free field in $d$
   dimensions.

  As a second example we  consider a $V$ interaction of
  particles in the hyperbolic space (22) (whose analytic continuation $y\rightarrow iz$ gives de Sitter space).
   In such a case eq.(66) has the
  solution (at the real time $y\simeq z^{\frac{1}{d}}$)

  \begin{equation}
\phi^{in}(y, x)=y^{\omega}\int d p\Big( a(
  p)H_{\omega}^{(1)}(C\vert p\vert y)+ a^{+}(
  p)\overline{H_{\omega}^{(1)}}(C\vert p\vert y)
  \Big)\exp(i px)
\end{equation}
where $\omega=\frac{d}{2}$.

 If the interaction is of the form $\delta(y)V$
   then the propagators are logarithmic and the
model will be superrenormalizable for (normal ordered) polynomial
$V$. There will be no ultraviolet divergencies in the S-matrix.

\section{Summary}
We find it as a remarkable property of some singular second order
differential operators in $D$ dimensions that the Green function
restricted to the boundary can be less singular than $\vert
x-y\vert^{-D+2}$. When we put  an interaction on the boundary then
the regularity of the model can be extended to models with a
non-trivial scattering. Models with an interaction on the boundary
are certainly of physical relevance in statistical physics. It is
still unclear whether such models are acceptable in high energy
physics as , e.g., the brane theories. We suggest however that
restricting an interaction to a brane can give a promising way of
avoiding ultraviolet divergencies in quantum field theory. Our
construction of the quantum field theory with a boundary ( which
preserves the symmetries of the boundary) is different than the
AdS-CFT approach. However, there appear some interesting relations
between both approaches if we treat the boundary value $\phi_{0}$
as random (quantum) and average over $\phi_{0}$.

\end{document}